\documentstyle[11pt,newpasp,twoside,epsf]{article}
\def\etal{{\rm et al.\,}}
\def\kms{{\rm km\,s$^{-1}$\,}}

\def\edcomment#1{\iffalse\marginpar{\raggedright\sl#1\/}\else\relax\fi}
\reversemarginpar

\begin{document}
\title{New Nitrogen and Carbon in AF-supergiants}
 \author{Kim A. Venn}
\affil{Macalester College, 1600 Grand Avenue, Saint Paul, MN, 55105, and
University of Minnesota, 116 Church Street, Minneapolis, MN, 55455 }
\author{Norbert Przybilla }
\affil{Institute for Astronomy, 2680 Woodlawn Drive, HI 96822}

\begin{abstract}
The AF-supergiants in the Galaxy and the SMC allow us to examine
predictions from evolution models through their CNO abundances.
In these proceedings, we recalculate the NLTE nitrogen abundances 
in 22 Galactic and 9 SMC A-supergiants using improved atomic data
and model atmospheres to compare with new evolution models. 
The new abundances are higher than previously published values, 
and suggest that most of these stars have undergone substantial 
mixing with CN-cycled gas.   While there is no clear relationship
with mass, there is an apparent relation with metallicity since
the SMC stars (including B-stars) have larger nitrogen enrichments.
We suggest that rotational mixing is indicated from the main-sequence
throughout the supergiant range, with more substantial rotational
mixing in the SMC stars.   In addition, the SMC AF-supergiants appear 
to have undergone the first dredge-up during a previous red giant phase,
and possibly the Galactic AF-supergiants have as well.   All abundances 
are compared to the new solar abundances from M. Asplund (this conference).
\end{abstract}

\section{Introduction}
Reviews on massive star evolution have shown there are a number
of observations that suggest an additional parameter affects the
evolution scenarios, ranging
from blue-to-red supergiant ratios in clusters to CNO abundances 
in main-sequence and evolved stars; c.f., Maeder \& Meynet (2000),
Heger \& Langer (2000 = HL00), Maeder \& Conti (1994).   
In Galactic supergiants, the N/C ratios have long supported a
parameter that varies from star-to-star, like rotation.   This
is because a variety of N/C ratios have been found within clusters
of stars (thus, not natal variations), and because the N/C ratios
do not clearly scale with effective temperature, luminosity, or mass.

Evolution models that include effects of rotation emerged in 2000.
Not only can rotation help to explain the abundance ratios and
certain mass discrepancies, but the new models also predict that
massive stars may produce and mix primary nitrogen into their 
atmospheres (by Meynet \& Maeder 2002).    The new models have
been calibrated based on abundance ratios, masses, rotation rates,
and ages in the published literature.  But they also make definate
predictions that can be tested for further refinement.   

Unfortunately, one direct test that is difficult to implement is to 
examine abundance ratios versus rotation rate.   Rarely do we know the
inclination angle of a rotating star.   Also, as $v$sin$i$ increases,
then the spectral lines become so shallow and broad that the atmospheric
analysis becomes quite difficult and less certain.   Fortunately, there
are other indirect tests.

\section{Boron and Rotation}

One of the new predictions from stellar evolution calculations that
include rotation is that the surface abundances of lithium, beryllium,
and boron (LiBeB) will be rapidly depleted before significant mixing of 
any other material (see Heger \& Langer 2000).    
These elements are fragile and easily destroyed when exposed to hot protons,
thus any mixing in the atmosphere will rapidly dilute LiBeB and destroy 
those atoms that are mixed downwards. 
This should be detectable in main-sequence stars; but boron is the 
only element with measurable absorption lines in B-type stars.

\begin{figure}
\plotfiddle{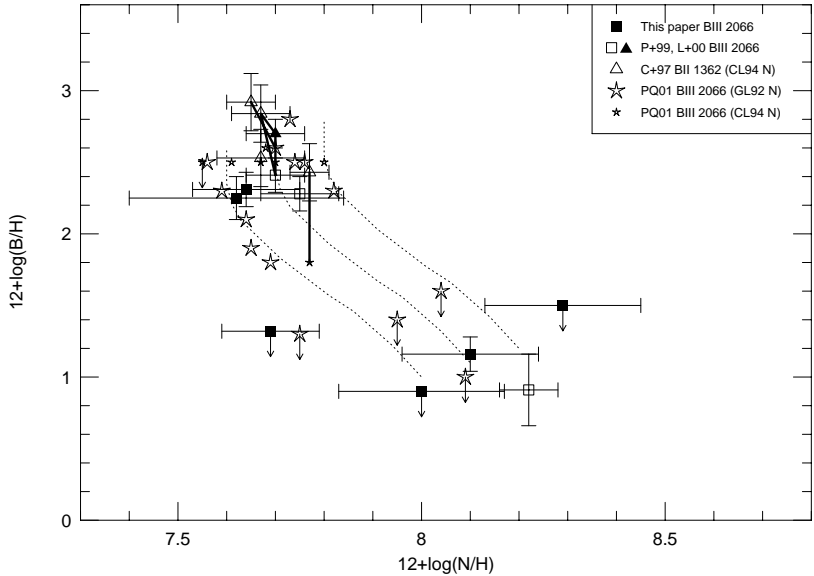}{8cm}{0}{45}{45}{-180}{-30}
\caption{Boron verus Nitrogen in main-sequence B-type stars
from Venn \etal 2002.   Notice the extreme range in boron
abundances observed at the same low nitrogen abundances. } 
\end{figure}

Boron has been measured in a handful of main-sequence B-stars 
(see Figure~1) from the BIII 2066 and BII 1362 resonance lines.   
Both require UV spectral observations, such as with HST-STIS.
There is a large range in the initial boron abundances in 
main-sequence B-stars, by a factor of 10 to 100.    This variation is
not natal since several stars within a single association show this
range, for example, the star in the Orion association. 

In Figure~1, boron is observed to be depleted {\it before} there 
is any detectable nitrogen enrichment.   This matches predictions, 
as seen from the HL00 model for a 12 M$_\odot$ star at 200 \kms 
through H-core burning and three sets of initial abundances 
[(B, N) = (2.6, 7.6), (2.8, 7.7), (2.8, 7.8)]. 
Thick lines connect boron abundances for the same stars from different 
analyses.   The boron abundance papers are listed in the legend 
(see references), the nitrogen abundances are from Cunha \& Lambert 
(1994 =CL94), Gies \& Lambert (1992 =GL92), or Vrancken \etal (2000). 
Currently, the most extreme rotating models suggest that boron can 
be depleted by a factor of 10 before nitrogen is enriched 
(e.g., M = 25 M$_\odot$, v$_{rot} \ge$ 400 \kms).   The current data
(especially star HD\,36591 in Ori OB1) suggest even larger depletions
are possible before nitrogen enrichment.  

These boron depletions in main-sequence B-star, 
{\it without significant nitrogen enhancements}, 
are unique to rotating stellar evolution models. 
It is possibly to reproduce a boron-depletion {\it with} 
a nitrogen-enhancement through mass-transfer in a binary system
(Wellstein, Langer, \& Braun 2001).

\section{CNO in Intermediate-Mass Supergiants}

Another prediction of the rotating models is that there should be
a wide range in the C and N abundances in evolved massive stars if
these stars had a range in rotational velocities on the main-sequence.
This has been seen and is well documented for Galactic stars in the 
literature, e.g., in O-stars (Herrero \etal 1992) in B-supergiants
(McErlean \etal 1999, Lennon \etal 1988) and in 
A-F supergiants (Venn 1995).    More recently, Smartt \etal (2002)
found that Sher 25, an evolved early B-supergiant surrounded by 
nebular ejecta, has only a small nitrogen enrichment, incompatable 
with the predicted first dredge-up abundances. 
The fact
that many of the N/C ratios in these stars ranged from solar ratios to 
only slightly enriched supports the idea that it is rotation that 
affects these abundances.   Another possibility has been that the N/C
ratios are affected by surface mixing and the first dredge-up during 
a previous red giant phase;  in this case, we expect an offset in the
mean N/C ratio (of $\sim$+0.5~dex for stars $\ge$10 M$_\odot$), which
has not been seen in the literature until now (below).

In most of the supergiants, non-LTE corrections have had to be 
applied to the C, and especially the N, abundances.    The non-LTE
corrections result in N/C ratios in Galactic supergiants that show
no relationships with effective temperature, luminosity, or mass.    
And a variety of N/C ratios are often found for stars in the same
OB association.  Thus, the variations in N/C are best attributed to 
star-to-star variations, such as rotation.   

{\bf New Solar Abundances:}
Before proceeding with a comparison of N/C ratios, it is important
to notice that the solar abundances have been significantly improved
through the 3-D convection models by M. Asplund and collaborators.
In this conference proceedings, new solar CNO abundances are presented,
showing excellent agreement between a variety of abundance indicators
for each element.   In the rest of this paper, we will adopt these
new solar abundances; 12 + log (C, N, O, Fe) = (8.41, 7.80, 8.66, 7.45). 

\begin{table}
\caption{Galactic Nitrogen Abundances: 12 + log({\it N/H})$_{\rm NLTE}$ }
\begin{tabular}{lll|lll}
\tableline
HD & NEW & OLD &    HD & NEW & OLD \\
\tableline
87737  & 8.54 $\pm$0.03 (3) & 8.09 $\pm$0.06 &   34578  & 8.33 $\pm$0.04 (3)  & 7.97 $\pm$0.14  \\
46300  & 8.28 $\pm$0.14 (8) & 7.85 $\pm$0.10 &   147084 & 8.44 $\pm$0.10 (3)  & 8.23 $\pm$0.12  \\ 
161695 & 8.48 $\pm$0.11 (8) & 8.03 $\pm$0.11 &   58585  & 8.26 $\pm$0.16 (3)  & 8.16 $\pm$0.16  \\
195324 & 8.67 $\pm$0.02 (4) & 8.14 $\pm$0.10 &   222275 & 8.51 $\pm$0.03 (2)  & 8.31 $\pm$0.25  \\
207263 & 8.43 $\pm$0.05 (3) & 8.04 $\pm$0.04 &   67456  & 8.19 $\pm$0.02 (2)  & 8.06 $\pm$0.22  \\
175687 & 8.26 $\pm$0.01 (3) & 8.29 $\pm$0.07 &   36673  & 8.19 (1)            & 8.30 $\pm$0.13  \\ 
3940   & 8.15 $\pm$0.02 (3) & 7.75 $\pm$0.11 &   148743 & 8.19 $\pm$0.13 (3)  & 7.99 $\pm$0.19  \\
14489  & 8.01 (1) & 8.01                     &   6130   & 8.00 $\pm$0.14 (9)  & 7.98 $\pm$0.12  \\
13476  & 8.34 $\pm$0.06 (2) & 7.96 $\pm$0.08 &   25291  & 7.96 $\pm$0.11 (7)  & 7.85 $\pm$0.11  \\
15316  & 8.35 $\pm$0.03 (3) & 7.96 $\pm$0.12 &   196379 & 7.61 $\pm$0.20 (11) & 7.57 $\pm$0.20  \\
210221 & 8.63 $\pm$0.20 (2) & 8.19 $\pm$0.04 &   59612  & 8.57 $\pm$0.02 (2)  & 8.29 $\pm$0.12  \\
\tableline
\end{tabular}
\end{table}

\section{New Nitrogen Abundances}

Much of the sample for the N/C ratios in Galactic supergiants comes 
from the A-F supergiants (Venn 1995).   Since then, new collisional
excitation data has become available for NI (Frost \etal 1998), which does
affect the non-LTE corrections (Przybilla 2002, and this proceedings), 
particularly for the hotter stars.    This atomic data
was sorely needed for the most reliable nitrogen abundance determinations
since the collision excitation cross-sections were the most significant
source of uncertainty in the Lemke \& Venn (1996) nitrogen model
calculations in A-stars.  Now, the new atomic data shows that the 
NI level populations are more collisionally coupled, which reduces the 
NLTE {\it corrections} (which increases the NI abundances).

In this conference proceedings, we present a new table of nitrogen 
abundances for the A-F supergiants; see Table~1.   These new results
include the updates to the corrections calculated by NP 
(see Przybilla 2002 for a detailed description), 
which makes a substantial difference to the hotter supergiants, yielding 
higher nitrogen abundances.   Only the NI $\lambda\lambda$7440 and 
$\lambda\lambda$8700 multiplets, and $\lambda$8629 line are included in
these new NLTE calculations.  The new NLTE corrections do not significantly
affect the cooler stars since their NI atomic levels are already more closely 
collisionally coupled.  Thus, similar calculations to those in Venn (1995) 
have been used for the six coolest stars (which include more lines).
We have also recalculated the ATLAS9 model atmospheres
using the most recent unix version.
This has only a minor effect on the final abundances. 
The old abundances from Venn (1995) are also shown for comparison.
The average nitrogen abundance has significantly increased, and the
distribution in abundances with effective temperature is shown in
Figure~2. {\it Note that Asplund's solar abundances listed above
are used in these new figures}.

In Figure~2, there is no clear trend in the N abundances with temperature
above 8000~K.    All stars show similar large N enrichments.   The coolest 
stars show smaller N enrichments though, with one star (HD196379) showing
no indication of internal mixing.
We also show the NLTE CI abundances from Venn (1995), 
as well as new NLTE CII abundances in three stars from Przybilla (2002). 
The carbon abundances in the cool stars are from
the weak $\lambda$7115 multiplet and show very small NLTE corrections.
However the carbon abundances in the hotter supergiants (hollow circles)
are from the strong $\lambda$9100 multiplet which have very large, and
less certain NLTE corrections.   In general, the N enrichments are
accompanied by smaller C depletions, as expected in CN-cycled gas.
Finally, the N/C ratios are shown.    The N/C ratios scatter about
the predicted first dredge-up abundances (on the red giant branch)
for 9 to 20 M$_\odot$ stars from the non-rotating evolution
tracks of Schaller \etal (1992). 

\begin{figure}
\plotfiddle{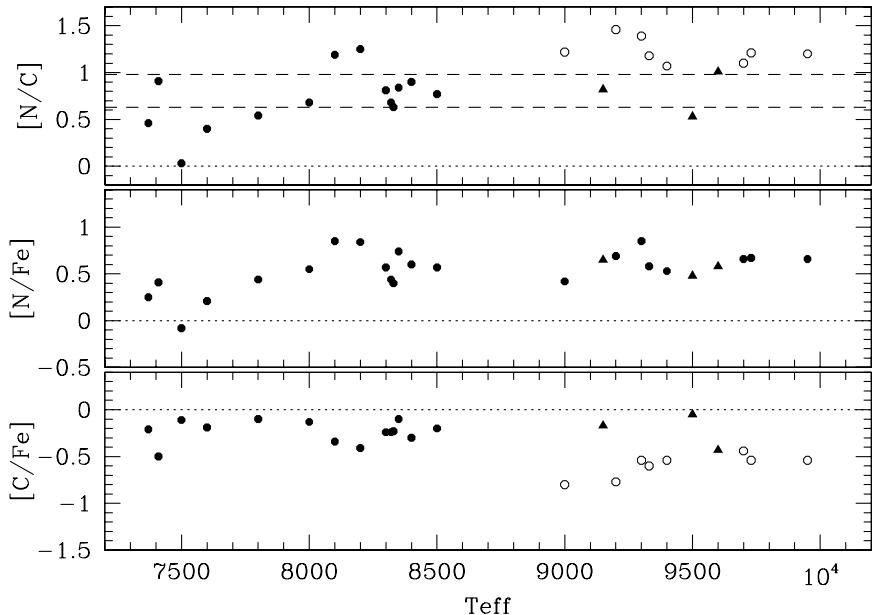}{8cm}{0}{60}{60}{-165}{-20}
\caption{The NLTE C and new NLTE N abundances in AF-supergiants.  
For C, filled circles include the CI $\lambda$7115 results from Venn (1995), 
and hollow circles represent the more uncertain CI $\lambda$9100 results 
(and their corresponding N/C ratios). 
Filled triangles are the NI and CII results from Przybilla (2002).   
Dashed lines show first dredge-up predictions for 9 to 20 M$_\odot$ stars
above solar (the dotted line).
} 
\end{figure}

\section{SMC Abundances}

We have also recalculated the NLTE NI abundances in the SMC AF-supergiants
from Venn (1999).   Metallicity has a negligible effect on the
statistical equilibrium of nitrogen in these atmospheres.  The new SMC 
NI abundances are listed in Table~2 and plotted in Figure~3.  Nearly all 
stars have higher nitrogen abundances from the new calculations.
For only the two coolest stars, where the levels are more closely
collisionally coupled, have we used similar calculations from 
Venn (1999, which also include more lines). 
Like with the Galactic stars, the new NLTE nitrogen abundances
are higher than the published values, and there continues to be
no clear trend in the abundances with temperature.    Also like
the Galactic stars, one very cool star shows no significant nitrogen 
enrichment.   
{\it Notice that the {\rm initial} [N/Fe] ratio for SMC stars is significantly
below solar (dotted line) because of the different chemical evolution 
of this galaxy}.

\begin{table}
\caption{SMC Nitrogen Abundances: 12 + log({\it N/H})$_{\rm NLTE}$}
\begin{tabular}{lll|lll}
\tableline
AzV & NEW & OLD & AzV & NEW & OLD \\
\tableline
110 & 7.84 $\pm$0.06 (2) & 7.64 $\pm$0.18  &   254 & 7.28 (1)           & 7.13 $\pm$0.27 \\
298 & 7.72 $\pm$0.07 (2) & 7.65 $\pm$0.21  &   442 & 7.55 $\pm$0.20 (3) & 7.32 $\pm$0.21 \\
136 & 7.73 $\pm$0.05 (3) & 7.26 $\pm$0.07  &   213 & 7.79 $\pm$0.18 (4) & 7.74 $\pm$0.22 \\
463 & 7.07 (1)           & 6.92 $\pm$0.25  &   174 & 6.79 (1)           & 6.76 \\
478 & 7.87 $\pm$0.07 (3) & 7.61 $\pm$0.16  & \\
\tableline
\end{tabular}
\end{table}

\subsection{Primary Nitrogen?}
That the SMC nitrogen abundances are so
much larger than the first dredge-up predictions (indicated by the
solid line in Figure~3) suggests these stars have undergone excessive 
mixing.   Even if these are post-RGB stars, 
their nitrogen indicates even more mixing as
predicted by the new rotating evolution models.   Does this mean
that there is primary nitrogen in the atmospheres of these metal-poor
supergiants (as predicted for metal-poor stars by Meynet \& Maeder 2002)?
At present, we can only say that the metal-poor SMC stars have more
nitrogen in their atmospheres.   Its nucleosynthetic history is an open
question.

\section{More Rotation?  First Dredge-Up?}

A comparison of the N abundances in Galactic B-stars, Galactic
AF-supergiants, and SMC AF-supergiants is shown in the histograms
in Figure~4.   Clearly the new Galactic AF-supergiant results 
are offset from the initial (solar and B-stars) N value by $\sim$0.5~dex.
This was not seen in the previously published results (dashed line, 
where the mean nitrogen abundance was close to the initial N abundance).  
In the SMC, the same broad distribution in nitrogen is seen as before, 
but now with more stars at higher values.

\begin{figure}[b]
\plotfiddle{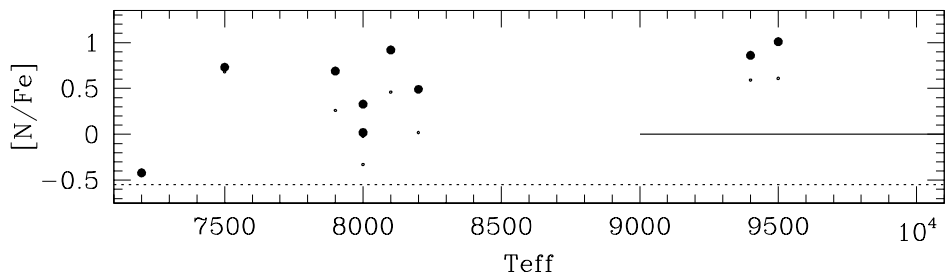}{3.1cm}{0}{65}{65}{-185}{-20}
\caption{New NLTE NI abundances in SMC AF-supergiants. 
Old results from Venn 1999 are shown by small dots.
The initial [N/Fe] ratio indicated by dotted line.   First dredge-up 
predictions by a solid line.} 
\end{figure}

In Maeder \& Meynet's review (2000), their Figure~6 shows that
a large spread in the N abundances in the A-F supergiants are
expected through variations in the rotational velocities.   This 
spread can be larger in higher mass stars or lower metallicity
stars.   We certainly see a larger spread in the SMC stars versus
the Galactic stars, and we expect this is due to metallicity since
the stars have similar mass ranges (see Figure~5).

\begin{figure}
\plotfiddle{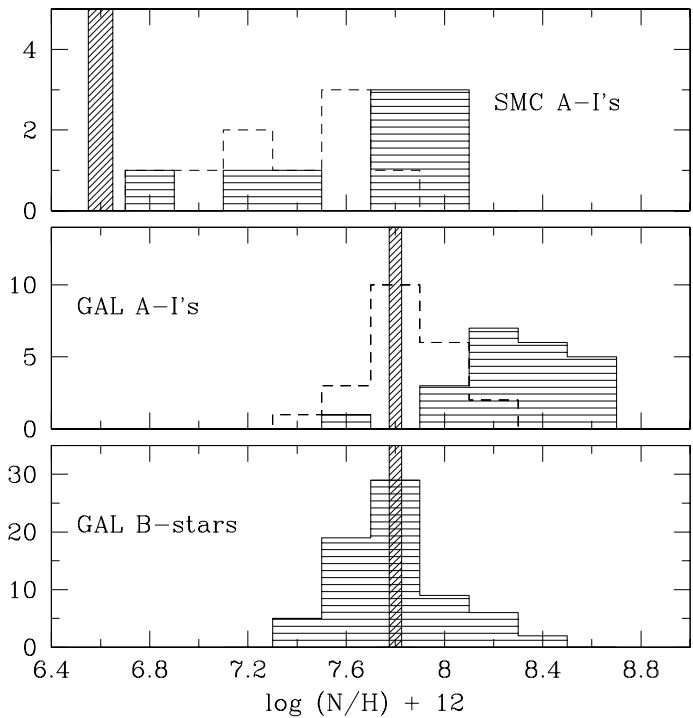}{9.2cm}{0}{50}{50}{-150}{-10}
\caption{Histogram of NLTE nitrogen abundances in Galactic B-stars
(GL92, CL94), and Galactic and SMC A-type supergiants.   The old
histogram results from Venn (1999) are indicated by the dashed line.
The initial solar and SMC ISM nitrogen abundances are indicated. 
} 
\end{figure}

Since all of the AF-supergiants exhibit significantly enriched 
N abundances, and the metal-poor SMC stars show the largest 
enrichments of all, then these results suggest:

(a) All of the stars had very high rotation rates ($>$300 \kms)
on the main-sequence.   This does not seem likely since the 
samples include alot of stars and the mean $v$sin$i$ values for 
these stars when on the main-sequence (B2-type) is only $\le$200 \kms 
(e.g., Fukuda 1982, de\,Jager 1980). 

(b) The stars have higher masses than expected.   
This is possible, but unlikely.  The stars appear to range 
in mass from 5 to 25 M$_\odot$ based on a comparison 
with standard evolution tracks (Figure~5), and even if those 
tracks are not appropriate (e.g., rotational mixing can affect
the post-main sequence evolution) there does not appear to be a relationship 
in the N or N/C ratios with mass.  

(c) Rotational mixing is more efficient than currently predicted.

(d) The stars have undergone the first dredge-up during a previous
red giant phase.   This certainly supports the N and N/C ratios 
observed, but since rotational mixing must also happen on the main-sequence
(to explain the boron dispersion in main-sequence B-stars), then 
the scatter in the N abundances and N/C ratios could be attributed
to early rotational evolution effects in addition to first dredge-up
abundances.   In the SMC, additional mixing beyond the first dredge-up 
predictions is clearly indicated. 
[Of course, the few coolest stars with low N (SMC) or N/C ratios (Galactic)
could not have undergone the first dredge-up yet and must be evolving 
directly from the main-sequence]. 

\begin{figure}
\plotfiddle{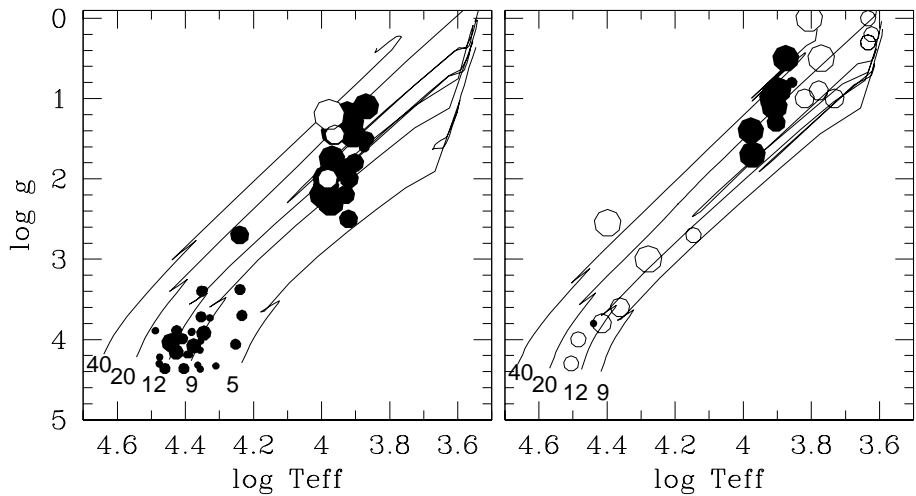}{6.5cm}{0}{65}{65}{-180}{-10}
\caption{N and N/C ratios for B-stars and supergiants in the
Galaxy (left panel) and the SMC (right panel).  The size of the 
point corresponds to the [N/C] ratio (Galactic) or [N/H] overabundance
(SMC).   Hollow circles in the left panel are Przybilla's (2002) N/C 
ratios.  Galactic B-star abundances from CL94, GL92. 
SMC B-star results from R+02 (including AV304, small solid point), 
L+96, R+93, L+91, R+90.  
SMC FG supergiant results from H97, B+91, LL92.
} 
\end{figure}

Either of the last two options is likely given the current
theoretical models.   In addition, post-RGB evolution is 
indicated for blue supergiants (from a He-burning sequence 
parallel to the main-sequence) in metal-poor Local Group 
dwarf irregular galaxies from deep and accurate HST 
colour-magnitude diagrams   
(e.g., Dohm-Palmer \etal 1997 (SexA), 1998 (Gr8); 
Tolstoy \etal 1998 (LeoA)).
Thus, excessive N enrichments in the SMC AF-supergiants
are most simply explained as post-RGB first dredge-up abundances.   
But if we also consider the range in the nitrogen abundances, and
the SMC B-star nitrogen enrichments, then main-sequence rotational 
mixing, as well as RGB convective mixing, are both indicated in 
metal-poor stars. 

Is the first dredge-up indicated at Galactic metallicities?
The new AF-supergiant nitrogen abundances do cluster around the
first dredge-up predictions.   But we also know now that rotational
mixing is important in interpreting abundance ratios, in SMC
supergiants and in Galactic main-sequence stars (e.g., boron). 
Thus, the Galactic AF-supergiants may also have been RGB stars
in their past, but it is also possible that their [N/C] ratios
could be explained by excessive rotational mixing alone.

\end{document}